\documentclass[a4paper,prb,twocolumn,showpacs,preprintnumbers,amsmath,amssymb,groupeaddress,superscriptaddress]{revtex4}

\usepackage{graphicx}
\usepackage{dcolumn}
\usepackage{bm}
\usepackage{amsmath}

\newcommand{\VEV}[1]{\left\langle {#1}\right\rangle} 

\def\SS{\mbox{\boldmath $S$}}
\def\H{\mbox{\boldmath $H$}}
\def\B{\mbox{\boldmath $B$}}

\def\LSMO{La$_{2-2x}$Sr$_{1+2x}$\-Mn$_2$O$_7$}
\def\LSMOc{La$_{1-x}$Sr$_{x}$\-MnO$_3$}
\def\PLSMO{(La$_{1-z}$Pr$_z$)$_{1.2}$Sr$_{1.8}$Mn$_2$O$_7$}
\def\uPLSMO{La$_{1.2}$Sr$_{1.8}$Mn$_2$O$_7$}

\newcommand{\commentout}[1]{}

\def\addrParma{Dipartimento di Fisica e Unit\`a CNISM, Universit\`a degli Studi di Parma, Viale G.\ Usberti 7A I-43100 Parma, Italy}
\def\addrDresden{Institute for Solid State Research, IFW Dresden, 
P.O.\ Box 27\,01\,16 D-01171 Dresden, Germany}
\def\addrParis{Laboratoire de Physico-Chimie de l'Etat Solide,
 UMR 8648 B\^at. 414 Universit\'e Paris-Sud, 91405 Orsay, France}
\def\addrRomania{Department of Physical, Theoretical and Materials Chemistry, 
Faculty of Chemistry, "Al. I. Cuza" University, Carol I, 700506 Iasi, Romania}

\begin{document}

\title{Magnetic order in double-layer manganites 
(La$_{1-z}$Pr$_z$)$_{1.2}$Sr$_{1.8}$Mn$_2$O$_7$: intrinsic properties and role of the intergrowths.}

\author{G. Allodi}
\email{allodi@fis.unipr.it}
\author{M. Bimbi}
\author{R. De Renzi}
\affiliation{\addrParma}
\author{C. Baumann}
\affiliation{\addrDresden}
\author{M. Apostu}
\affiliation{\addrParis}
\affiliation{\addrRomania}
\author{R. Suryanarayanan}
\author{A. Revcolevschi}
\affiliation{\addrParis}
\date{\today}

\begin{abstract}

We report on an investigation of the double-layer manganite series 
\PLSMO\ ($0\!\le\! z\!\le\! 1$), 
carried out on single crystals by means of both macroscopic magnetometry and 
local probes of magnetism ($\mu$SR, $^{55}$Mn NMR).
Muons and NMR demonstrate an antiferromagnetically ordered
ground state at non-ferromagnetic compositions ($z \ge 0.6$),
while more moderate Pr substitutions 
($0.2\le z\le 0.4$) induce a spin reorientation transition within the 
ferromagnetic phase.
 
A large magnetic susceptibility is detected at 
$T_{c,N} \!<\! T \!< \!250$~K at all compositions. From $^{55}$Mn NMR 
spectroscopy, such a 
response is unambiguously assigned to the intergrowth of a ferromagnetic 
pseudocubic phase (La$_{1-z}$Pr$_z$)$_{1-x}$Sr$_{x}$MnO$_3$, with an overall 
volume fraction estimated as $0.5-0.7$\% from magnetometry.
Evidence is provided for the coupling of the magnetic moments of 
these inclusions with the magnetic moments of 
the surrounding \PLSMO\ phase, as in the case of finely dispersed impurities.
We argue that the ubiquitous intergrowth phase may play a role in the marked 
first-order character of the magnetic transition and the 
metamagnetic properties above $T_c$ reported for double-layer manganites.

\end{abstract}


\maketitle
\section{Introduction}

In the last decade, double layer strontium manganites \LSMO\ have attracted 
the interest of the colossal magnetoresistance (CMR) community due to a 
magnetoresistive effect even larger than in the pseudocubic compounds 
\LSMOc, and for their peculiar anisotropic magneto-transport properties.  
These systems are the $n=2$ members of the Ruddlesden-Popper (R-P) series 
$La_{n(1-x)}$Sr$_{1+nx}$Mn$_{n}$O$_{3n+1}$,
constituted by blocks of two adjacent magnetic MnO$_{2}$ 
planes, decoupled from each other by a rocksalt-like block of non-magnetic 
oxide (La,Sr)O.
At optimum doping ($x=0.4$), the material is a quasi 2-dimensional 
ferromagnetic metal (FMM) with a reported Curie temperature 
$T_c \approx 125$~K \cite{nature1996} and a highly anisotropic electrical 
conductivity $\sigma$ (two orders of magnitude larger in the $ab$ plane than 
along the $c$ axis). \cite{nature1996,science1996} 
In some sense, 
these compounds may be viewed as a natural kind of FMM-oxide multilayers, 
and hence prototype systems for the study of the physics of spin-polarized 
electrons in heterostructures, which is the clue for the production of 
spintronic devices.

As for their pseudo-cubic counterparts, the electronic and magnetic states 
of these materials result from the competition of different interactions, 
comparable in energy scale. While Zener's double exchange (DE) dominates 
in the FMM state ($0.32 \le x \le 0.45$), 
electronic correlations drive the system through a number of different 
insulating phases with distinct types of magnetic, charge, and orbital order 
as a function of temperature and hole doping $x$, leading to a very complex 
phase diagram. \cite{mitchell2000} 
Because of the reduced dimensionality, however, the DE term is less effective 
in the double-layered compounds, as indicated by 
their strongly diminished Curie temperatures
($T_c = 125$~K and 370~K in \LSMO\ and \LSMOc, respectively, 
around optimal doping $x=0.33$).
Further DE energy reduction may arise from cation size mismatch when different
alkali-earth metals and rare earths are substituted. 
The small ionic radius of Pr$^{3+}$ induces distortions and tilting of the 
MnO$_6$ octahedra, which depress the hopping term
both by directly reducing the overlap of the Mn an O orbitals, 
and by altering the relative populations of the 
$e_g$ orbitals. \cite{apostu_prl2003}
As a result, the FMM phase may become unstable, and the systems exhibit 
metal-insulator (MI) transitions.

The isoelectronic series \PLSMO\ shows indeed an insulating  
non-ferromagnetic ground state for $z>0.5$. Nevertheless, the FMM phase still 
exists at these compositions as a metastable one, into which 
the system can be driven by 
applying a magnetic field of order 
a few Tesla at low temperature, or by cooling the sample in an even smaller 
applied field. \cite{apostu_prl2004} The former process is actually a 
field-induced first order MI 
transition, and the corresponding drop in electrical resistivity may be viewed 
as another kind of CMR. The system displays peculiar thermal and 
magnetic hysteresis effects, and the FMM state survives even after removal of 
the magnetic field. \cite{apostu_prb2001}

While the MI transition has been subject of vast investigation and is well 
understood, \cite{apostu_prb2004,apostu_prb2005} magnetism in the non-FMM 
phases of this series has been debated 
for long time. No spontaneous magnetic order has 
ever been demonstrated at $z>0.5$ in the so-called virgin state 
(zero field, zero field cooled) at low temperature, and hence the systems have 
been assigned to a spin-glass state. \cite{apostu_prl2003}
Moreover, the nature of the paramagnetic phase itself in double-layer 
manganites is the subject of a controversy reaching well beyond the issue of 
La-Pr substitution. These systems exhibit in fact enhanced magnetic 
susceptibility well above the highest known $T_c$ of a double-layer manganite, 
up to close to room temperature. 
At the same temperatures, ARPES recently showed evidence 
in La$_{1.24}$Sr$_{1.76}$Mn$_2$O$_7$ for a locally metallic phase in a 
globally insulating state. \cite{arpes}
Although the possible presence of spurious  
FMM inclusions in the sample has been considered in many papers, 
there is however general disagreement about their role in the anomalous 
responses detected in the paramagnetic phase. According to some authors,
such responses are essentially intrinsic spin and electronic properties of the 
double-layer material, \cite{apostu_epl,apostu_prb1999, arpes}
while they are mostly due to the intergrowth of a secondary
FMM phase for some other. \cite{bader}

In this paper, we address the issue of the magnetic state of La-Pr 
double-layer manganites by means of local probes of magnetism, namely,
muon spin rotation ($\mu$SR) and $^{55}$Mn NMR. Implanted 
muons are sensitive probes of local 
magnetic fields in the matter, suited to investigate the magnetic structure 
even in the case of short-range 
order or dilute magnetic clusters, where probes in the reciprocal space cannot 
detect any magnetic Bragg peak. In addition, 
$^{55}$Mn NMR in zero external field provides a determination of 
the spontaneous hyperfine fields at $^{55}$Mn nuclei, which constitutes a 
probe of both magnetic order and the electronic state of the Mn ions. 
The investigation is complemented by magnetization and magnetic susceptibility 
measurements. By combining these techniques, we demonstrate 
that the $z>0.5$ compositions  
display short range antiferromagnetic order 
below a N\'eel temperature 
$T_N \approx 60$~K, while any ferromagnetic response above that temperature is 
to be assigned to pseudocubic FMM inclusions, of order 1\% of sample volume.  

The paper is organized as follows. 
Section \ref{sec:experimental} provides a brief description of sample 
preparation and the experimental methods, as well as the essential guidelines 
to the NMR technique applied to the study of manganites. 
In section \ref{ssec:results.musr} we describe the $\mu$SR results on the 
changes in magnetic structure across the phase diagram. In section 
\ref{ssec:results.magn} we present evidence
for ferromagnetic intergrowths from magnetometry and susceptometry. 
The electronic nature of the intergrowths and their peculiar coupling to the 
bilayer manganite domains are further elucidated by $^{55}$Mn NMR in section 
\ref{ssec:results.nmr}.
Results are discussed in section \ref{sec:discussion}.

\section{Experimental}
\label{sec:experimental}
\subsection{Samples} 
\label{ssec:samples}
The single crystals were 
grown with the floating-zone method by using an image furnace, melting 
sintered rods of polycrystalline materials of the same nominal 
compositions obtained by standard
solid-state reaction. For $^{55}$Mn NMR on unsubstituted \uPLSMO\ at 
low-temperature, however, we employed the original
powder material instead of the final single crystal due to the small 
penetration depth of radiofrequency (rf) in a metal, making a bulk 
conductive sample opaque to the driving rf field. 
A subset of these crystals have been subject to the investigations reported in 
Refs.\ \onlinecite{apostu_prb2001,apostu_prb2003,apostu_epl,apostu_prb2004,apostu_prb2005}. 
Preliminary magnetic characterization by macroscopic dc magnetization 
measurements in a field of 10~mT confirmed 
ferromagnetic order at compositions $z \le 0.4$ with 
Curie temperatures decreasing from $\approx 110~$K in the Pr-free compound, 
down to $Tc \approx 60~$K at $z=0.4$, and a non-FM state at $z\ge 0.6$. 
\cite{apostu_prb2003}

\subsection{Magnetometry}
\label{ssec:exp.magnetometry}
Measurements were carried out by means of a Quantum Design 
MPMS-XL SQUID device, employed either as a dc magnetometer or as an ac 
susceptometer in zero bias field. Magnetic ac susceptibility was also measured
vs.\ temperature in the rf domain from the increase in inductance of a 
small coil ($L_0 \approx 200$ nH in air) wound around the sample and placed in 
a liquid nitrogen cryostat. The coil was driven 
by a rf field of $\approx 0.01$~Oe at a frequency of $\approx 50$ MHz
by a Hewlett-Packard 4191A impedance meter, to which it was connected
by a $\lambda / 2$ coaxial cable for the working frequency.
The inductance $L$ is related to 
the rf volume magnetic susceptibility $\chi_{rf}$ by
\begin{equation} 
(L-L_0)/L_0 = f4\pi\chi_{rf}/(1+N4\pi\chi_{rf})
\label{eq:chirf}
\end{equation} 
where N and f are the 
demagnetization and filling factors of the sample, respectively. In the case
of a bulk conductive sample, a significant diamagnetic contribution to 
$\chi_{rf}$ may also arise from eddy currents. This rf method is therefore 
intended as a qualitative characterization, only sensitive to large FM 
components.

\subsection{$^{55}$Mn NMR in manganites}
\label{ssec:exp.nmr}
Nuclear magnetic resonance experiments were performed by means of a 
home-built phase-coherent spectrometer, \cite{hyrespect} using an 
Oxford EXA field-swept cold-bore cryomagnet as a sample environment. 
Spectra were recorded point by point at frequency steps of 
1 MHz by a standard $P-\tau-P$ spin echo sequence, with delays $\tau$ of 3-5 
$\mu s$, and equal rf pulses $P$ of 0.5-2 $\mu s$ with intensity optimized 
for maximum signal. NMR signals from 
antiferromagnetic (AF) phases, characterized by a small rf enhancement 
factors (see below), were detected by a tuned probehead, while the strongly 
enhanced signals from ferromagnetic (FM) phases could be detected by a 
non-resonant circuit (a small coil terminated onto 50~$\Omega$) with a  
fully automated procedure. 

The $^{55}$Mn nuclear probe is coupled to the electronic Mn moments by the 
hyperfine interaction, of the form 
\begin{equation}
{\cal H} = \hbar ^{55}\gamma g\mu_B {\bm{S}}\cdot{\cal  A}\cdot\bm{I}
\label{eq:hf}
\end{equation}
giving rise to a spontaneous magnetic field 
$\bm{B}_{hf} = g\mu_B{\cal A}\VEV{\bm{S}}$ at the nuclei in the magnetically 
ordered state of the material. Here ${\bm S}, {\bm I}$ are the electron and 
nuclear spin, $g$ is the Land\'e factor, $\mu_B$ is the Bohr magneton, 
$^{55}\gamma$ is the nuclear gyromagnetic ratio and 
${\cal A}$ is the on-site hyperfine coupling tensor. All other terms,
such as transferred contributions and dipolar fields from nearest neighbor 
ions, have been neglected in Eq.~\ref{eq:hf} since they are much smaller.
 
From Eq.\ \ref{eq:hf}, 
the spontaneous resonance frequency, proportional to 
the hyperfine field, essentially depends on the on-site electronic spin, 
as well as on the valence of the magnetic 
ions and the metallic vs.\ insulating electronic state.
Experimentally, the isotropic component of the hyperfine coupling constant 
(the Fermi contact term) exhibits a value ${\cal A}_{iso} 
\approx 10$ T/$\mu_B$ throughout the iron transition metal series, 
approximately independent of the ionic species. \cite{RadoSuhl}
In the case of manganites, ${\cal A}_{iso}$ would give rise to
resonance frequencies of approx.\ 300 and 400 MHz for 
Mn$^{4+}$ ($S\!=\!3/2$) and Mn$^{3+}$ ($S\!=\!2$), respectively. 

The hyperfine field of Mn$^{3+}$, however, also exhibits a sizable 
pseudo-dipolar component 
$ g\mu_B {\cal  A}_{dip}\cdot\bm{S}$ of order 100 kOe, while  
the anisotropic coupling tensor
${\cal A}_{dip}$ vanishes for the 
octahedral symmetric Mn$^{4+}$ ion. 
As a consequence of this anisotropic 
coupling term the resonance frequency of $^{55}$Mn in Mn$^{3+}$ may lie 
anywhere in the 250-500 MHz frequency range, depending of the orientation of 
the electronic spin relative to the ionic symmetry axis. \cite{spinel}
In the presence of many magnetically nonequivalent sites, the resonance of 
$^{55}$Mn in Mn$^{3+}$ may be split into a number of satellite 
lines, as it was actually found in insulating low-doped \LSMOc. \cite{kapusta}
At increased structural or magnetic disorder, ${\cal A}_{dip}$ of Mn$^{3+}$ 
may simply contribute to a huge inhomogeneous linewidth, while the resonance 
of $^{55}$Mn in Mn$^{4+}$ are comparatively much sharper. 
In the metallic state of pseudo-cubic \LSMOc, however, individual 3+ and 4+ 
ions are replaced by a single Mn mixed valence of $3+x$, while 
the pseudo-dipolar coupling term is averaged out by fast electronic motion.
As a consequence, the broad and complex NMR spectrum typical of the insulating 
compositions collapses into a single motionally-narrowed line in the metallic
 state. 

Information about the type of magnetic order, ferro- or antiferromagnetic,
can be obtained by NMR from the so-called radiofrequency enhancement, 
consisting in an amplification of the effective rf field 
at the nucleus, due to the hyperfine coupling 
between nuclear and electronic spins. 
An applied rf field couples to the magnetization of a magnetic material 
and tilts its electronic moments, thus inducing a modulation of the hyperfine 
field at the nucleus at the same frequency. \cite{turov,portis}
Since the hyperfine field is of order several Tesla, a small rf perturbation 
usually gives rise to a large modulation of $B_{hf}$ at the same frequency. 
Hence, if $H_1$ is the rf field applied in the sample coil,  
the effective rf field driving the resonance at the nucleus is enhanced by a 
factor $\eta > 1$, $H_1^{eff} = \eta H_1$. 
As compared to NMR of the same nucleus in a non-magnetic 
materials, nuclear resonance can therefore be excited by reduced rf power. 
The same amplification mechanism holds for the NMR signal induced in
the pick up coil: $A \propto \eta N$, where $A$ is the signal amplitude, 
and $N$ is the number of resonating nuclei. 

The rf enhancement depends on the local magnetic structure and on the 
effective magnetic anisotropy. Clearly, $\eta$ is larger in ferromagnets 
because of their larger net magnetization coupling to the external rf field.
Two enhancement mechanisms may be distinguished in a ferromagnet: the applied 
rf field may displace the domain 
walls, or it may rotate the magnetization in the 
bulk of domains. The former comes into play for nuclei within domain 
walls and is generally the dominant mechanism in zero field, where it may give 
rise to values $\eta \approx 10^3-10^4$, while it vanishes in an applied 
static field above the saturation value. The latter is usually less effective 
by one-two orders of magnitude, and it is present 
both in zero and in an applied field. In special cases 
(e.g.\ uniaxial anisotropy and $H_1$ perpendicular to the easy 
axis), straightforward calculation yields 
a domain enhancement factor 
$\eta_d \propto H_{anis}^{-1}$, where $H_{anis}$ is the anisotropy 
field, while such a relation may only be taken as a qualitative one 
for the domain-wall enhancement $\eta_w$, whose actual dependence 
is much more complicated.
\subsection{$\mu$SR experiments}
\label{ssec:exp.musr}
The experiments were performed at Paul Scherrer Institut 
(Villigen, CH) on the GPS spectrometer in the spin-rotator (SR) setup. 
The asymmetry of the muon decay, $A(t)$, is directly proportional to the time dependent muon spin projection along the axis joining a detector pair at the opposite sides of the sample, which reveals the spin precession around the local 
field. \cite{Schenck} With the SR the 
initial muon polarization is rotated from the beam direction ($\hat x$ axis) 
by $\approx 47^\circ$ towards the vertical axis ($\hat y$), while emitted 
positrons were detected by two detector pairs along $\hat x$ and $\hat y$.
The crystals were mounted with the $c$ axis parallel either to $\hat x$ or 
$\hat y$, as more appropriate. This allowed 
simultaneous detection of the muon spin projections parallel and 
perpendicular to $c$.

In transition metal oxides muons stop at few (typically one or two) distinct interstitial sites. The muon spin is coupled to the electronic Mn moments by dipolar and hyperfine interactions, providing a different local magnetic field ${\bm H}_{\mu\,j}$ at each distinct site. Depending on its orientation each local field gives rise to precessing and to longitudinal components of the muon spin, revealed respectively by damped oscillations and purely relaxing terms in the muon asymmetry.  The muon polarization $P_\mu(t)$ may be fitted as
\begin{eqnarray}
P_\mu(t) = \frac {A(t)}{A_0} = \Big [ \mskip -22mu & & a_1 e^{-t/T_{1}} +  \nonumber \\
& & \sum_{j}a_{2j} e^{-t/T_{2j}} \cos \omega_j t\Big ]
\label{eqn:asym}
\end{eqnarray}
where $A_0$ is the total calibrated asymmetry for the GPS SR setup, $\omega_j=\gamma_mu H_{\mu\,j}$ are the muon precession frequencies, $\gamma_\mu/2\pi$ = 135.54 MHz/T is
the muon gyromagnetic ratio, $a_1, T_1$ and $a_{2j}, T_{2j}$ are the muon amplitudes and their respective relaxation times for the longitudinal and precessing components, respectively. Since precessing components are distinguished by their frequencies, whereas longitudinal components are not, only a cumulative term may be determined experimentally for the latter.

\section{Experimental results}
\label{sec:results} 
We report separately the results obtained by the different techniques, starting from $\mu$SR, then magnetometry and finally NMR.   
\subsection{$\mu$SR results}
\label{ssec:results.musr}
\begin{figure}
\includegraphics[width=\columnwidth]{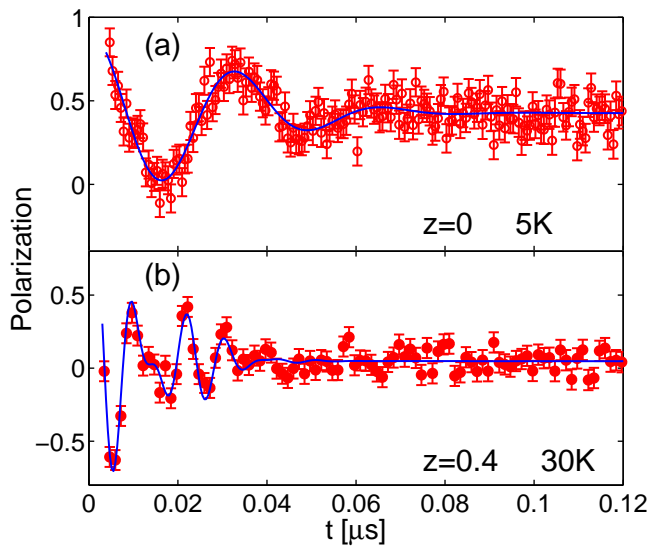}
\caption{\label{fig:precess}(Color online) Time evolution of 
muon spin polarization along the $c$ axis in a)   
La$_{1.2}$Sr$_{1.8}$Mn$_2$O$_7$ and b) 
(La$_{0.6}$Pr$_{0.4}$)$_{1.2}$Sr$_{1.8}$Mn$_2$O$_7$ at $T\ll T_c$.
}
\end{figure}

Precessing patterns are detected in the ordered phase at all FMM compositions, up to $z = 0.4$. Fig.\ \ref{fig:precess} shows the low temperature time dependent polarization at the two extremes of this range, with a best fit to Eq.~\ref{eqn:asym}. The analysis of frequencies and fractions for the different compositions, summarized below, reveal two distinct long range order magnetic structures at these two extremes, and a third form of short range magnetic order for higher Pr content. Relaxations in the paramagnetic phase, described separately, identify an additional contribution for all samples.

\begin{figure}
\includegraphics[width=\columnwidth]{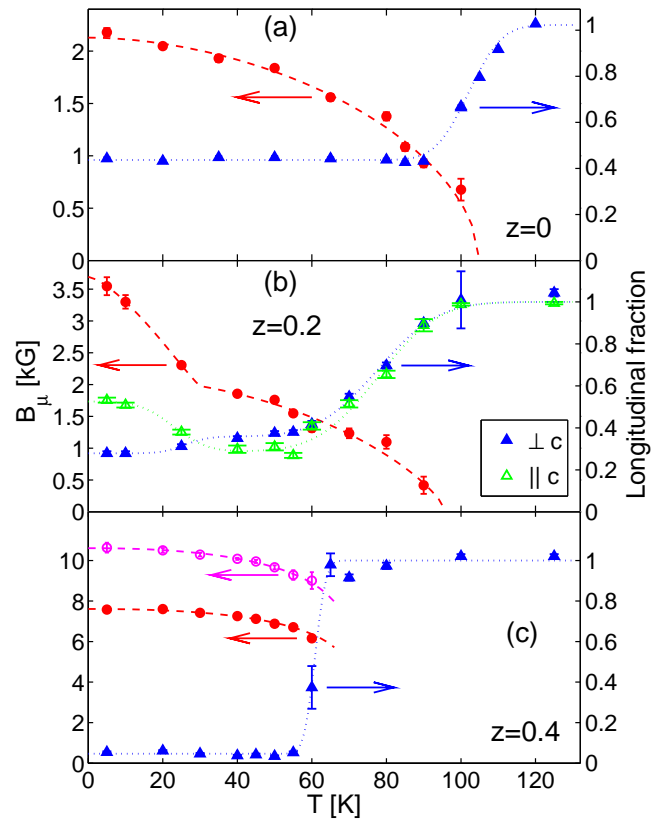}
\caption{\label{fig:musrFM}(Color online) Muon spontaneous fields (bullets) and longitudinal
 amplitudes (triangles) as a function of temperature, 
in the $z\!=\!0$ 
(panel a), $z\!=\!0.2$ (b), and $z\!=\!0.4$ (c) samples. Open and filled
triangles denote amplitudes along and perpendicular to the $c$ axis, 
respectively. Dashed lines are guides to the eye (dotted lines, see text).}
\end{figure}

\underline{$z=0$:} In the unsubstituted compound we detected a single precession 
component (Fig.\ \ref{fig:precess}a) with an internal field of 220 mT 
(29.6 MHz in frequency units) at $T=5$K, in close 
agreement with the value reported by Coldea {\it et al.} \cite{blundell} for 
the same nominal material. Such a field is
significantly smaller 
than the typical values reported for pseudocubic \cite{heffner,cestelli} and 
single-layer \cite{tiffel} manganites (of order 6-10 kG).
The precession frequency and the 
longitudinal amplitudes $a_1$ for the detector pair perpendicular to the $c$ 
axis are  plotted vs.\ temperature in 
Fig.~\ref{fig:musrFM}a. 
In the magnetically ordered state, we detected a finite longitudinal amplitude also for the spin projection $\parallel c$ (data not shown), indicating that the direction of  
the internal field ${\bm H}_\mu$ is tilted relative to the crystal axes.

The recovery of the full longitudinal amplitude, denoting the transition to the paramagnetic phase, may be fitted by a Gaussian distribution of Curie temperatures with a rather low average $T_c=105\pm 3$~K and width $\Delta T_c=8$~K (dotted line in Fig.~\ref{fig:musrFM}a). Although the distribution of transition temperature indicates a residual inhomogeneity, the internal field $H_\mu$, vanishes following the order parameter as $(T-T_c)^\beta$, as in regular second order transitions. 
This behaviour is markedly different from any previously 
reported in the literature for these metallic compositions. For instance Ref.~\onlinecite{shiotani} on a nominally similar La$_{1.2}$Sr$_{1.8}$Mn$_2$O$_7$, determines $T_c=126$K and a ``truncated'',
first-order-like transition, common to all  metallic 
manganites (with the exception 
of the large-bandwidth compound \LSMOc).
In such transitions the order parameter remains finite for $T \to T_c$, 
while the volume of the ordered phase shrinks as the transition temperature 
is approached. \cite{lynn, gubkin}

\underline{$z=0.4$:} Two precessing components 
could be resolved in the case of (La$_{0.6}$Pr$_{0.4}$)$_{1.2}$Sr$_{1.8}$Mn$_2$O$_7$ (Fig.\ \ref{fig:precess}b), with much higher internal fields 
of 7.6 and 10.6 kG at 
5~K (Fig.\ \ref{fig:musrFM}c). Their total precessing amplitude at $T\ll T_c$ 
is close to 100\% for the detector pair 
perpendicular to $c$ axis and it is negligibly small for that $\parallel c$, 
while the opposite occurs for 
the longitudinal amplitude. This demonstrates that the internal 
fields at both muon sites are collinear to $c$ within the accuracy of sample
orientation ($\pm 5^\circ$). 

This markedly different behavior of both  
 precession amplitudes and frequencies with respect to the 
unsubstituted compound arises from the spin reorientation induced by La-Pr 
substitution, reported by Vasil'ev {\it et al.}, 
\cite{jetpletters} who detect an easy magnetization axis perpendicular to $c$ 
in praseodymium-free \LSMO\ 
and parallel to the $c$ axis at $z\!=\!0.4$. 
Such a reorientation strongly affects the magnitude of the local field due to 
the anisotropy of the muon-electron interaction (essentially the dipolar 
coupling between the muon and electronic spins). 
The temperature dependence of the longitudinal amplitude 
(Fig.~\ref{fig:musrFM}c) reveals a rather sharp magnetic transition at 
$T_c=60\pm 1$~K. In contrast to the $z=0$ case this transition shows a strong 
first-order character, as witnessed by the high value of the internal fields 
at both muon sites just before the transition, 
$H_{\mu j}(T_c)/H_{\mu j}(0)\approx 0.85$.

\underline{$z=0.2$:} In the ordered phase of the intermediate compound 
(La$_{0.8}$Pr$_{0.2}$)$_{1.2}$Sr$_{1.8}$Mn$_2$O$_7$, 
muons precess at the same single frequency as in the unsubstituted material,
except at very low temperature (Fig.\ \ref{fig:musrFM}b). 
In the 30-60 K interval, the $\mu$SR spectra of this  
and the $z\!=\!0$ samples are closely similar, in particular the values of 
$H_\mu(T)$ and the relative weights of the longitudinal 
and transverse amplitude $A_1/A_2$ coincide within experimental errors.
In the $z\!=\!0.2$ sample, however, the recovery of the full longitudinal 
asymmetry takes place continuously on warming from 60 to 90~K, 
which indicates an even broader distribution of $T_c$. 
Moreover, the precession signal shows a clear anomaly at $T< 20$~K, 
consisting in an upturn of both the internal field and the precession 
amplitude perpendicular to $c$, along with a complementary reduction of the 
longitudinal amplitude for the same detector pair. The precession  
linewidth increases as well at $T\to 0$, 
which might mask the presence of two distinct though poorly resolved 
precession frequencies, as in the case of $z\!=\!0.4$. 
Following the discussion above, 
such a low-temperature dependence 
of the magnitude and direction 
of the local field ${\bm H}_\mu$ is indicative of a partial spin reorientation,
apparently governed by an energy scale of order a few tens of Kelvin at this 
composition.

\begin{figure}
\includegraphics[width=\columnwidth]{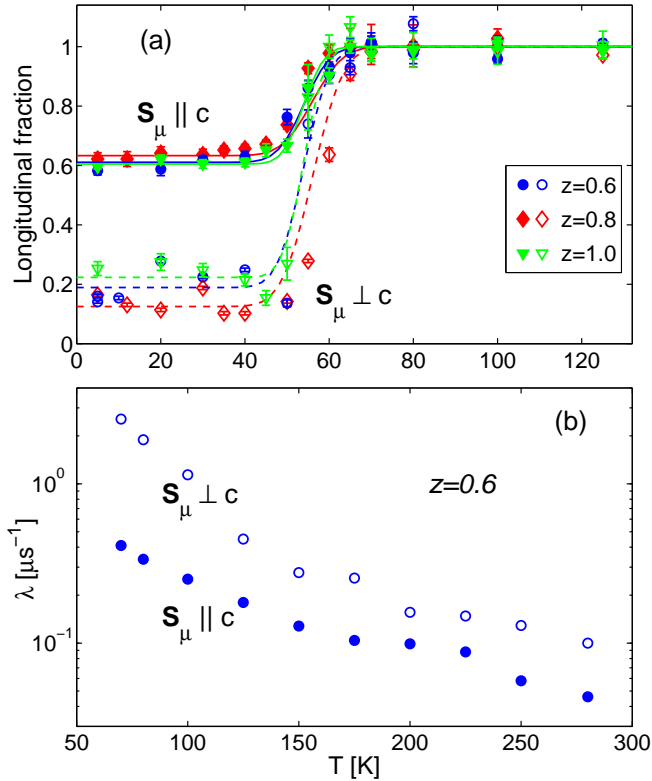}
\caption{\label{fig:musrAF}(Color online) a) Muon longitudinal amplitudes as a function of 
$T$ at AF compositions, parallel (filled symbols) and perpendicular 
to $c$-axis (open symbols). b) Longitudinal relaxations in 
(La$_{0.4}$Pr$_{0.6}$)$_{1.2}$Sr$_{1.8}$Mn$_2$O$_7$ at $T>T_N$ for the two 
muon spin components.
}
\end{figure}

\underline{$z\ge 0.6$:} The appearance of large static 
internal fields in the non-FMM compounds, originating from a non-fluctuating
component of the electronic spins, is proved by the drop of the longitudinal 
amplitudes below a magnetic transition temperature $T_N = 55\pm 5$\,K, nearly
independent of $z$ (Fig.\ \ref{fig:musrAF}a).
No spontaneous muon precession signal could be detected at these compositions 
after zero-field cooling. This indicates a very broad distribution of large
internal fields leading to 
an exceedingly fast depolarization of the transverse muon 
signal, which is already relaxed at early times.
In fig.\ \ref{fig:musrAF}a, comparison between the longitudinal signals 
detected parallel and perpendicular to the $c$ axis reveals however that the 
local field at the muon site ${\bm H}_\mu$,
though not collinear to the crystal axes, 
has a dominant component along $c$. Such a correlation of the internal field 
direction with the crystal lattice rules out a spin glass state and, 
in conjunction with the small macroscopic moment, \cite{apostu_prb2001} 
it demonstrates a 
low-temperature antiferromagnetic  order in the samples, although 
of short-range character. 

\begin{figure}[t]
\includegraphics[width=\columnwidth]{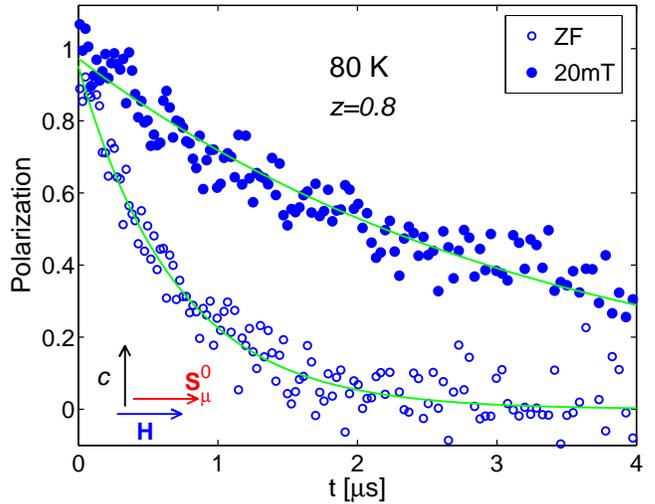}
\caption{\label{fig:quench08}(Color online) 
(La$_{0.2}$Pr$_{0.8}$)$_{1.2}$Sr$_{1.8}$Mn$_2$O$_7$: 
 muon spin polarization vs.\ time at 80~K for initial muon spin $\SS_\mu^0$ perpendicular to the $c$ axis, in zero field (open symbols) and external field  
$\mu_\circ H=20$~mT, $\H \parallel \SS_\mu^0$  (filled symbols).
}
\end{figure}

\underline{Relaxations:} The zero-field polarization of 
all the crystals in our series
exhibits exponential relaxations with sizable rates in the paramagnetic (PM) 
state, up to temperatures of order 200\,K, well above $T_{c,N}$.  We refer to 
these rates by the symbol $\lambda$, to explicitly avoid labeling them as 
longitudinal ($T_1^{-1}$) of transverse ($T_2^{-1}$).
Their magnitude, $\lambda \approx 1\mu\hbox{\rm s}^{-1}$ at 100~K, 
clearly prove their electronic origin, since  
random dipolar fields by static nuclei determine Gaussian relaxation
with much smaller rates, never exceeding $\approx 0.1\mu\hbox{\rm s}^{-1}$ in 
any known manganite compound. \cite{fanesi,tiffel} Strong dynamic relaxations 
in the paramagnetic regime may indeed originate from the critical slowing down 
of fluctuations, but they are usually confined to a narrow temperature 
interval just above $T_{c,N}$, contrary to our observations.

Figure \ref{fig:musrAF}b shows the temperature dependence of  $\lambda$ measured on a representative sample ($z=0.6$) in zero external field (ZF), in the PM phase. Two different rates are measured simultaneously in the SR setup (see Sec.~\ref{ssec:exp.musr}), equivalent to independent measurements in longitudinal detectors (i.e.~{\em along} the initial muon spin direction  $\SS_\mu^0$) with crystal c-axis aligned respectively parallel and perpendicular to the detector axis, hence to  $\SS_\mu^0$.
A marked relaxation anisotropy is apparent, with the 
ZF depolarization rate of the muon spin component perpendicular to the $c$ 
axis exceeding that of the parallel component by 
nearly one order of magnitude, up to well above $4T_N$. 
Also the unsubstituted crystal shows a qualitatively similar behavior above 
its transition temperature, albeit with slightly reduced rates.

If this unusual ZF relaxation were due to static magnetic moments, the value 
of the rate, $0.4\le \lambda \le 3\, \mu$s$^{-1}$, would be directly 
proportional to the static internal local fields, $0.4\le B_{\mu,s}\le 4 $ mT. 
Conversely, similar rates could be due to much larger instantaneous internal 
fields $B_{\mu}\gg B_{\mu,s}$, fluctuating with reciprocal correlation times 
$\tau^{-1}$ larger than their Larmor frequencies $2\pi\gamma_\mu B_\mu$ 
(dynamical rates scale with $B_\mu^2\tau$). In order to clarify this issue we 
measured the muon polarization in a magnetic field $\mu_0H > B_{\mu,s}$, which 
is expected to influence the rates in the first case, but not in the second.   

The time dependent polarizations in zero and in a longitudinal field of 20~mT
are compared in Fig.~\ref{fig:quench08} for another sample ($z=0.8$, but the 
behavior is generic) at 80\,K~$>T_N$ and with the initial muon 
spin $\SS_\mu^0$ perpendicular to the $c$ axis.
It is apparent that the application of a moderate longitudinal field 
(i.e.\ parallel to both $\SS_\mu^0$ and the detectors) {\em quenches} 
significantly the strong ZF polarization decay observed in this geometry. 
This clearly identifies the presence of a dominant static contribution at 
$T>T_{c,N}$, i.e.\ these zero field relaxations are essentially 
due to incoherent precessions in random static fields.
Only transverse static fields $\B_{\mu,s} \perp \SS_\mu^0$ contribute to such 
a mechanism; longitudinal ones, on the contrary, do not depolarize the muon 
spin. Therefore, the anisotropy in the ZF decay rates, 
$\lambda_{ab} \gg \lambda_{c}$, witnesses an anisotropic distribution of 
$\B_{\mu,s}$ peaked along the $c$ axis.
We remark that similar anisotropic relaxations
were reported in double-layer manganites by Heffner {\it et al.},
\cite{heffner98} who however overlooked 
their static character.

\subsection{Magnetic response}
\label{ssec:results.magn}

\begin{figure}
\includegraphics[width=\columnwidth]{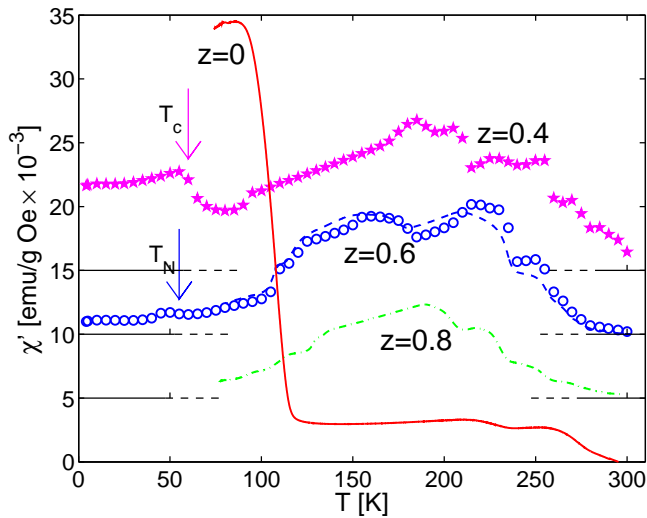}
\caption{\label{fig:chiac}(Color online) Ac susceptibility $\chi'$ ($H_{ac} \perp c$ axis) as 
a function of 
temperature, either measured by SQUID at acoustic frequency (markers)
or by the rf method described in the text (lines). 
For clarity, data are plotted with constant vertical offsets. 
}  
\end{figure}

We performed SQUID susceptibility measurements on the two samples 
closer to the MI phase boundary, 
namely $z=0.4$ and $z=0.6$, in an ac field of 0.1~Oe at 131~Hz applied in
the $ab$ plane, and dc field zeroed within a fraction of the earth magnetic 
field. 
In the $z=0.4$ sample, the onset of FM order below $T_c \approx 60$~K is 
marked by the rise of $\chi'$, while in $z=0.6$ the magnetic transition shows 
up as a shallow bump at $T_N\approx 55$~K (Fig.\ \ref{fig:chiac}).
Such low-temperature features, however, are 
overwhelmed at both FM and AF compositions by a huge response from 
270 down to 80~K, exhibiting a broad peak structure between 150 and 200~K, 
which drops on approaching $T_{c,N}$. 
This anomalous response in the PM phase is suppressed by a bias dc field of 
order few tens Oe, and static magnetization curves $M(T)$ in 
$H_{dc}\ge 100$~Oe (not shown) closely reproduced those published in Ref.\ 
\onlinecite{apostu_prb2003}.

Radio-frequency measurements in the 77-300~K temperature interval and zero dc 
field yield $L(T)-L_0$ data proportional to $\chi_{ac}$ as measured by SQUID 
at acoustic frequencies. 
\footnote{Mass rf susceptibilities are calculated from 
eq.~\protect{\ref{eq:chirf}} by assuming filling factors $f$ of order 
0.3\,-\,0.5 and neglecting the demagnetization term $N4\pi\chi_{rf}$.
} 
In \uPLSMO, the rf data provided an independent 
estimate of the Curie point, which is determined as $T_c=105\pm 3$~K from the 
inflection point of $L(T)$, in agreement with our $\mu$SR data.
In this as well as in the substituted samples the rf response also displays 
anomalous paramagnetic peaks above $T_{c,N}$, equivalent to the SQUID ones. 
Such peaks are more intense in the Pr-substituted samples than in \uPLSMO, and 
they are suppressed 
by a small dc magnetic field like those in $\chi_{ac}$ at 131 Hz.

\begin{figure}
\includegraphics[width=\columnwidth]{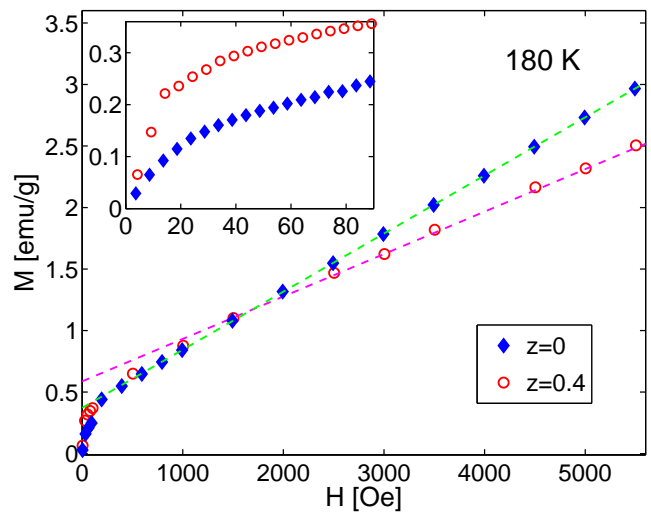}
\caption{\label{fig:m_h}(Color online) Magnetic moment vs. applied field ($H \perp c$ axis) 
at 180~K, 
at $z\!=\!0$ and $z\!=\!0.4$.
Inset: blow-up at small fields.
}  
\end{figure}

The insensitivity of the anomalous paramagnetic 
response to the driving frequency, up to the rf domain, demonstrates the
absence of significant dissipation and rules out, for instance, a spin glass
state. Indeed, dc magnetization measurements reveal a net moment in the 
PM phase which rather points to the presence of FM inclusions in the samples.
Fig.\ \ref{fig:m_h} plots typical $M(H)$ curves at $T> T_{c,N}$,
showing an initial steep growth of the magnetization $M$, followed by a much 
reduced  linear increase at higher fields. The change of slope takes place at 
a few tens Oe, roughly the same field value required to suppress the anomalous 
high temperature  contribution to $\chi'_{ac}$.
Clearly, the initial growth corresponds to the low field saturation of the 
magnetic moments of the inclusions, while the linear behavior at higher fields 
is the genuine PM response of the double-layer manganite. 
These FM impurities obviously coincide with the intergrowths 
suggested by $\mu$SR.
From the intercept of the linear high field behavior at $H=0$, 
$M_0\approx 0.4(1)$ and 0.6(1) emu/g respectively in the Pr-free and 
Pr-substituted compounds, we estimate the corresponding intergrowth fractions 
to be 0.5(1)\,\% and 0.7(1)\,\% of the crystal volume, respectively.

The small saturation field of these clusters,
along with the absence of detectable magnetic hysteresis,
demonstrate their soft magnetism at temperatures which are
well above the $T_{c,N}$ of the pure double layer manganite.
A drop of the ac susceptibility takes place below 120-150~K. This is evident in the AF members, where $\chi'(T)$ reaches a minimum 
just below $T_N$, it is still distinguishable in the $z=0.4$ sample, whereas it is masked by the double-layer ferromagnetic response in the $z=0$ sample.
The strong decrease of the high temperature $\chi'(T)$
indicates an increase of their magnetic 
anisotropy of the intergrowth on cooling, suggesting that the 
magnetic hardening of the FM clusters might be induced by the AF order in the 
host matrix via their exchange coupling with 
the double layer manganite.

\subsection{Zero-field $^{55}$Mn NMR}
\label{ssec:results.nmr}

\begin{figure}
\includegraphics[width=\columnwidth]{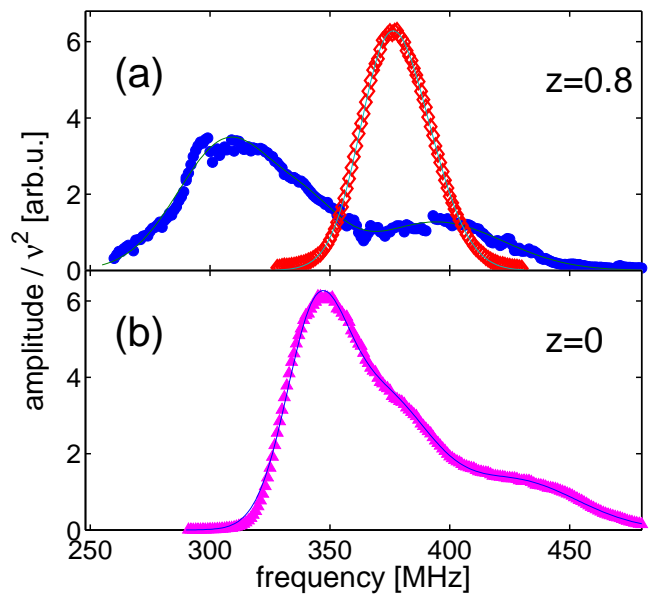}
\caption{\label{fig:NMRz0z08}(Color online) a) 
(La$_{0.2}$Pr$_{0.8}$)$_{1.2}$Sr$_{1.8}$Mn$_2$O$_7$: zero-field $^{55}$Mn NMR 
spectra at 1.6K of the proper (filled symbols) and impurity phases 
(open symbols). Plotted intensities are only corrected for the NMR spectral 
sensitivity $\propto\nu^2$.
 b) Zero-field $^{55}$Mn NMR spectrum of La$_{1.2}$Sr$_{1.8}$Mn$_2$O$_7$ at
5 K. }
\end{figure}

In the non-ferromagnetic samples $z\ge 0.6$, two distinct kinds of $^{55}$Mn 
NMR signals were detected in zero field (ZF-NMR) at 
low temperature, whose spectra are 
shown in Fig.\ \ref{fig:NMRz0z08}a for a representative sample ($z=0.8$). 
The narrower single peak at 377 MHz, characteristic of a pseudocubic DE phase, 
was excited with a sizable enhancement $\eta \approx 300$, slightly smaller 
than the typical values previously reported for the zero-field $^{55}$Mn 
signal in CMR manganites, \cite{noi,kapusta} while the broader two-peak 
spectrum in the 270-450 MHz range was detected with a much smaller 
enhancement value $\eta \approx 10$. The amplitudes integrated over 
the spectrum of the two signals are in the ratio $1 : 1.3$, 
respectively. Since however their enhancement factors are in a ratio 
$\approx 30$, the actual fraction of nuclei from which the former signal 
originates is estimated as 
$\approx 2\pm 1$\,\%, 
\footnote{The large uncertainty on the amplitude ratio
follows from the extrapolation of the spin echo amplitudes at $\tau=0$ in 
the presence of short $T_2$ (especially for the AF signal), as well as from
the inhomogeneity of $\eta$, which prevents simultaneous optimum excitation 
of all nuclei.} 
and therefore the single-peak spectrum is due 
to a minority phase. 

The minority phase is both ferromagnetic and metallic, as indicated by the 
large rf enhancement and its single-peak motionally narrowed spectrum, 
respectively. The proper spectrum of insulating \PLSMO\ is the broad two-peak 
one and the presence of a zero-field resonance signal, hence well defined 
 ($(\overline{\Delta B_{hf}^2})^{1/2}/\overline{B_{hf}}=\Delta\nu/\nu\approx 0.15$) static hyperfine fields at the nuclei, 
is by itself further independent 
evidence for a magnetically ordered ground state of this material. The
moderate value of $\eta$ observed for this signal is compatible with AF order 
in the presence of either a small canting, or significant exchange coupling 
with the FMM minority phase (see below).
All the insulating samples of our series ($z\ge 0.6$) display both kinds of 
signals, with very similar features.     

Fig.\ \ref{fig:NMRz0z08}b shows the $^{55}$Mn spectrum of 
La$_{1.2}$Sr$_{1.8}$Mn$_2$O$_7$ ($z=0$), characterized by a large rf 
enhancement $\eta$. More complex spectra were reported \cite{shimizu} for this 
composition; however  they were obtained on the bulk metallic single crystal, 
where eddy-current shielding might emphasize surface impurity phases. Our 
spectrum, detected on powders, consists again of two peaks
(a third one, less intense, fitted at $\approx 380(4)$~MHz, is compatible with 
the known amount of intergrowth in this sample), which are however narrower 
and shifted to higher frequency with respect to the intrinsic $z\ge 0.6$ 
spectrum (cfr.\ Fig.\ \ref{fig:NMRz0z08}a and b). The nuclear hyperfine 
spectroscopy at $z=0$ is definitely more complex than the isotropic-hyperfine, 
single peak detected for the time-averaged Mn valence, $3+x$, in all 
pseudocubic DE compounds, e.g.\ La$_{0.67}$Sr$_{0.33}$MnO$_3$ \cite{noi} or 
La$_{0.67}$Ca$_{0.33}$MnO$_3$. \cite{bibes} 
The reason for such a difference might rely on a residual orbital order 
in the quasi two-dimensional double-layer manganite, in contrast with the
orbital liquid state of the metallic pseudocubic material. 
A detailed interpretation in terms of the orbital and magnetic structure is
however outside the scope of this article.
 
\begin{figure}
\includegraphics[width=\columnwidth]{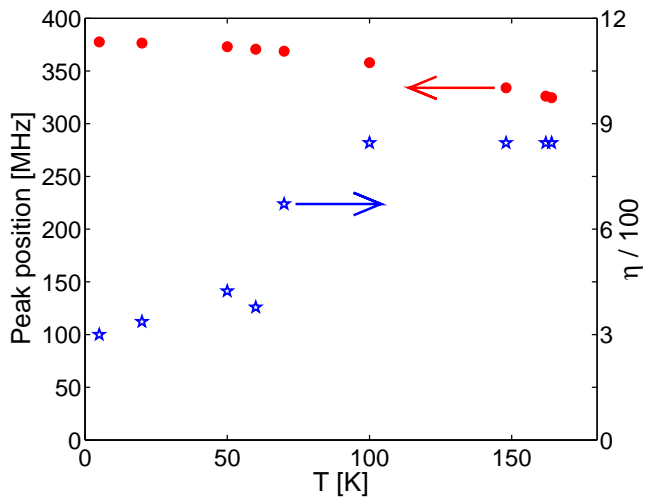}
\caption{\label{fig:NMRci}(Color online)  
(La$_{0.4}$Pr$_{0.6}$)$_{1.2}$Sr$_{1.8}$Mn$_2$O$_7$: mean resonance frequency 
(bullets) and rf enhancement (stars) of the impurity phase signal.  
}
\end{figure}

Here, we can simply conclude that the $^{55}$Mn spectra of the double-layer DE 
manganite, of the insulating double-layer manganite and of the FM intergrowth 
are clearly distinguishable, both by shape and enhancement. The latter in 
particular is very different from the former two and it identical to those 
recorded in the pseudocubic FMM phases. Its ZF-NMR resonance line could be 
followed up to approximately 170 K, above which
observation of the signal was hindered by a very short spin-spin relaxation 
time $T_2$. Its mean resonance frequency $\bar{\nu}_L$, 
proportional to the order parameter, 
is plotted vs.\ temperature in Fig.\ \ref{fig:NMRci} for the $z=0.6$ 
sample. 
At 165 K, this quantity shows only a 15\% reduction relative to its
low temperature value, indicating that the actual critical temperature of this 
impurity phase is significantly higher.
The presence of a ZF-NMR
signal well above $T_N$, in the same temperature range where both macroscopic 
magnetization and $\mu$SR detect an extrinsic response, demonstrates the 
coincidence of the impurity phase probed by NMR with the ferromagnetic 
clusters revealed above $T_N$ by the other techniques.
In addition, from the unique shape of their spectra, $^{55}$Mn NMR 
unambiguously assigns such ferromagnetic clusters to 
the intergrowth of metallic $n\gg 2$ member of the R-P series 
whose physical properties are probably indistinguishable from those of the 
pseudocubic $n=\infty$ member).

From Fig.\ \ref{fig:NMRci}, the order parameter 
$\bar{\nu}_L(T)$ of the impurity phase is insensitive to the N\'eel 
temperature of the double-layer material, whereas the rf enhancement factor 
$\eta(T)$ exhibits a step-like increase 
above $T_N\approx 60$ K by a factor of 3, which indicates 
a reduction of the effective anisotropy in the impurity phase, i.e. the 
magnetic softening of the ferromagnetic intergrowth at the paramagnetic 
transition of the host AF structure. This implies a sizable exchange coupling 
between the impurity and majority phases of 
the sample, in agreement with the large surface to volume fraction of the 
intergrowth, and it also justifies of the broad peak in the ac susceptibility 
$\chi'(T)$ above $T_N$.

\section{Discussion and conclusions}
\label{sec:discussion}

We detect an antiferromagnetic short-range-order ground state in the $z>0.5$ members of \PLSMO. The zero field 
$^{55}$Mn NMR signal reveals a 
 peak structure characteristic of the on-site hyperfine field and the small enhancement value appropriate for an insulating
antiferromagnet. The drop of the muon longitudinal amplitudes below $T_N$, 
and their anisotropy with respect to crystal orientation,
confirm a static magnetic order with the direction of the moments strongly correlated to the crystal $c$ axis. The AF structure is further supported by the absence of a macroscopic magnetic moment and the short range nature is deduced from the lack of a coherent $\mu$SR precession, in this context a sign of strong inhomogeneous relaxations.

The known spin-reorientation \cite{jetpletters} along the $c$ axis for $z=0.4$ 
is witnessed by our muon data in all $z\ge 0.4$ samples. Since the same 
phenomenon is observed both in the present case, where the La-Pr substitution 
reduces the metallic bandwidth, and in Pr-free samples, towards the border of 
the metallic FM phase, \cite{mitchell2001} where it is caused by a hole 
concentration becoming critically low, it looks as a generic feature connected 
to the weakening of the double-exchange interaction.
 
Our unsubstituted \uPLSMO\ crystal deserves special attention. 
This sample exhibits a Curie temperature of 105~K, rather lower than any 
previously reported $T_c$ for this material (e.g.\ $T_c\approx 125$~K 
according to Refs.\ \onlinecite{mitchell2000,kubota,shiotani}). 
On the other hand, the spontaneous field at the muon of 2.2~kG detected at 
low-temperature is in close agreement with the value reported in the 
literature for this material. \cite{blundell} Such a comparatively low field 
value corresponds to FM order with an easy magnetization axis in the $ab$ 
plane. 
Electronic moments oriented along the $c$ axis (as in $z=0.4$) or AF ordered, 
\cite{blundell} on the contrary, would give rise in fact to much higher 
precession frequencies.

Therefore, the discrepancy of the Curie point in our sample  
cannot be due to a deviation from its nominal stoichiometry. 
According to the phase diagram reported in the literature 
\cite{mitchell2000,kubota} the present value of $T_c$ would correspond to 
hole concentrations $x$ differing from the nominal value by 
more than 0.06, well beyond the preparation accuracy. 
Such large deviations in stoichiometry would also determine a significantly 
different magnetic structure: either canted AF, in the overdoped case,  
or FM with out-of-plane spin orientation, \cite{mitchell2001}
like for our (La$_{0.6}$Pr$_{0.4}$)$_{1.2}$Sr$_{1.8}$Mn$_2$O$_7$, in the 
opposite case. The former circumstance is incompatible with low temperature 
magnetization data, showing a saturation moment of $\approx 3.6$~$\mu_B$ per 
Mn ion as appropriate for collinear ferromagnetism. 
The latter
contrasts with the $\mu$SR spectrum, which is peculiar of a FM structure with 
an in-plane easy axis, as pointed out above, and hence it is a signature of a 
hole concentration close to the optimum value for maximum $T_c$. 
We therefore maintain that ours is a very good Pr-free specimen of $x=0.4$ 
stoichiometry.

\begin{figure}
\includegraphics[width=\columnwidth]{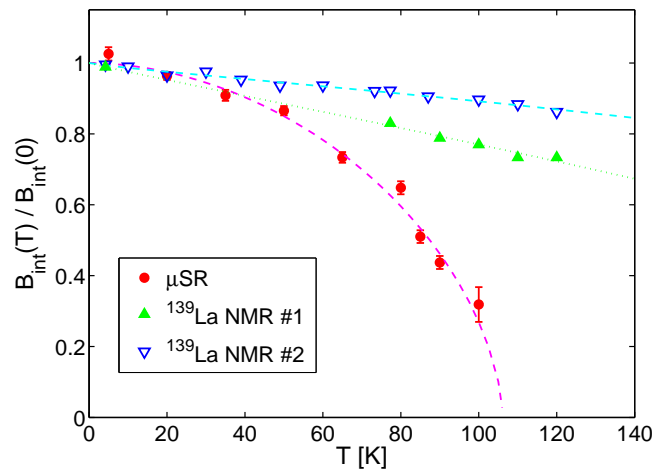}
\caption{\label{fig:secondorder}(Color online) \uPLSMO: normalized $\mu$SR precession 
frequency in the present single crystal (bullets), 
compared with the normalized spontaneous $^{139}$La NMR frequencies at 
the intra-bilayer site from the nominally equal compounds of 
Refs.\ \protect{\onlinecite{unpublished}} (\#1, filled triangles)
and \protect{\onlinecite{shiotani}} (\#2, open triangles), as a function of 
temperature. The reference $^{139}$La frequencies at $T=0$ are 29.9 and 
24.3 MHz for samples \#1 and \#2, respectively.
}
\end{figure}

Actually, the present \uPLSMO\ crystal displays a true second order phase 
transition characterized by a vanishing order parameter at the Curie point, as 
it appears from Fig.~\ref{fig:secondorder}, where quantities proportional to 
the magnetic order parameter are plotted versus temperature for different 
samples: the $\mu$SR Larmor frequency in the present 
sample with the $^{139}$La NMR spontaneous frequencies 
of two nominally identical compounds, the crystal studied by Shiotani 
{\it et al.}, \cite{shiotani} and a polycrystalline sample of ours 
prepared by standard solid-state reaction, \cite{unpublished} respectively. 
All frequencies are normalized to their zero temperature values. The order 
parameter of our crystal tends to zero for $T\to T_c$ as in a regular second 
order transition, whereas the other two curves remain very high close to their 
transition temperature. Other samples with higher $T_c$ were reported to 
exhibit magnetic transitions with strong first order character, 
\cite{berger_jap}
also accompanied by marked metamagnetic properties in the PM phase, namely, 
the recovery of FM order in applied fields of a few tesla at tens of degrees 
above $T_c$. \cite{shiotani,nature1996} 

The explanation for the remarkably different behavior of our single crystal at 
$T_c$ leads us to discuss the role of the intergrowths. Stacking faults
in the double layer structure with numbers of adjacent MnO$_2$ plane in excess
of two are common in all R-P crystals of transition metal oxides and they were 
actually observed in \uPLSMO\ by high resolution electron 
microscopy. \cite{bader}

SQUID characterization reveal an amount of cubic intergrowth five time smaller 
in this crystal than in our polycrystalline sample 
of Fig.\ \ref{fig:secondorder}, and nearly a factor two smaller than our other 
Pr-substituted crystals. 
We argue that the peculiar behavior of our crystal is due to its relative 
high purity. 
The double layer manganite is exchange coupled to the intergrowths, which are known to occur as stacking faults, extended in the (001) planes and very short in the c direction. This topology maximizes their effect on the magnetic free energy balance, possibly leading to a natural exchange-spring behavior, \cite{Asti} 
which could result in distinct magnetic properties of the intergrowth-rich samples. It is very likely therefore that the magnetic phase transition of pure 
\uPLSMO\ is second order and takes place at a Curie point not exceeding 
$\approx 105$~K, as in our sample, while higher $T_c$ values and 
first-order-like phase transitions are both extrinsic properties, driven by 
the exchange coupling with the intergrowth 
in specimens with larger fractions of such inclusions. 
The same exchange coupling would also account for a metamagnetic behavior above $T_c$, which we actually observed in our powder sample, but not in the crystal. If this scenario were confirmed, double layered manganites samples of
standard quality (i.e.\ impure ones) should be regarded as a natural kind of 
exchange spring materials, with new physical properties 
arising from the interaction between the two phases.

Magnetometry and$^{55}$Mn NMR provide independent and consistent evidence for 
the presence of intergrowths in our crystals, up to approximately 1\,\% 
volume in the Pr-substituted crystals, and to a few percent in our powder 
specimen.  
NMR demonstrates that these intergrowths display
a ferromagnetic metallic state very close to  
that of the pseudocubic bulk material. In addition,
NMR and susceptibility data reveal a peculiar increase of their magnetic 
anisotropy in coincidence with the  
magnetic ordering of the proper double layer manganite phase, 
more evident at AF compositions  
thanks to the magnetic contrast between the inclusions and the host matrix.
Such a magnetic hardening constitutes the proof of a strong 
exchange 
coupling between the two phases, which points to a large surface to volume 
ratio of the impurity domains.

Direct evidence of the FMM intergrowths comes also from the muon relaxations 
of Fig.~\ref{fig:musrAF}b and \ref{fig:quench08}, 
which demonstrate the static nature of the excess relaxation for 
$\bm{S}^0_\mu\perp\hat{c}$. Only the presence of a modest {\em static} local 
field $\bm{B}_\mu$ at the muon site may explain the quenching effect of 
Fig.~\ref{fig:quench08}. 
Such field cannot be understood in terms of typical paramagnetic properties, 
e.g. those of the \PLSMO\ matrix alone, whereas it is simply explained by 
distant diluted ferromagnetic impurities.
However the relaxations of Fig.~\ref{fig:musrAF}b indicate an unusually large 
anisotropy of $\bm{B}_\mu$ together with a pronounced increase of the rates 
for temperatures decreasing towards $T_c\approx 60$ K. Both findings are 
difficult to understand in terms of simple dipolar fields from distant diluted 
inclusions. At present we cannot offer a detailed explanation of these 
observations. We notice that they may  indicate that the \PLSMO\ matrix is not 
passive, in the presence of magnetized inclusions and that the two intermixed 
phases give rise to specific, coupled, peculiar magnetic properties.

In conclusion, we have demonstrated a magnetically ordered ground state in 
optimally hole-doped La-Pr double-layer manganites at all praseodymium 
concentrations, and the ubiquitous presence of nanoscopic metallic pseudocubic 
inclusions, seemingly unavoidable with the material synthesis, 
which undergo independent FM ordering at higher temperature.
Evidence for exchange coupling between the two phases suggests that the 
 physical properties of double-layer manganite may be strongly affected by 
the intergrowths. 
 
\section*{ACKNOWLEDGMENT}
The authors thank A. Amato for assistance and helpful
discussion. Partial funding by NANOFABER and OFSPIN, 
and the technical support of the LMU staff and of the
accelerator staff of the Paul Scherrer Institute are gratefully
acknowledged.

\end{document}